\documentclass[prb,aps,twocolumn]{revtex4}
\usepackage{amssymb}
\usepackage{amsmath}
\usepackage{graphicx}
\usepackage{epstopdf}
\begin{document}
\title{Dynamical Mean Field Study of the Two-Dimensional Disordered
Hubbard Model} \author{Yun Song$^{1,2}$, R. Wortis$^1$, W. A. Atkinson$^1$ }
\affiliation{${}^1$Department of Physics and Astronomy, Trent University,
1600 West Bank Dr., Peterborough ON, K9J 7B8, Canada\\
${}^2$Department of Physics, Beijing Normal University, Beijing 100875, China} 
\date{\today}
\begin{abstract}
We study the 
paramagnetic Anderson-Hubbard model 
using an extension of dynamical mean-field
theory (DMFT), known as statistical DMFT, that allows us to treat
disorder and strong electronic correlations on equal footing.  
An approximate nonlocal Green's function is found for individual disorder realizations and
then configuration-averaged.  We apply this method to two-dimensional
lattices with up to 1000 sites in the strong disorder limit, where an atomic-limit 
approximation is made for the self-energy.
We investigate the scaling of the inverse participation ratio at
quarter- and half-filling and find a nonmonotonic dependence of the
localization length on the interaction strength.  For strong disorder,
we do not find evidence for an insulator-metal transition, and the
disorder potential becomes unscreened near the Mott transition.
Furthermore, strong correlations suppress the Altshuler-Aronov density
of states anomaly near half-filling.
\end{abstract}
\pacs{71.10.Fd,71.27.+a,71.30.+h,73.20.Fz}
\maketitle

\section{Introduction}
The physical properties of interacting disordered
materials are often qualitatively different from their noninteracting
counterparts.  For example, it has long been known that noninteracting
quasiparticles in two-dimensional (2D) materials are localized by
arbitrarily weak disorder; however, there is evidence that
interactions can drive an insulator-metal transition in
2D.\cite{Kravchenko2004}  Similarly, there is a growing awareness that
disorder can fundamentally alter the physical properties of
interacting systems.  This arises in a number of transition-metal
oxides,\cite{Sarma1998,Nakatsuji2004,Kim2005,Kim2006} where the
predominantly d-orbital character of the conduction electrons results
in a large intra-orbital (on-site) Coulomb interaction relative to the
bandwidth.  These materials are of interest because they have an
interaction-driven insulating (Mott-insulating) phase and because of
the variety of exotic phases, such as high temperature
superconductivity, which appear near half-filling.  However,
transition metal oxides are typically doped by chemical substitution
and, with few exceptions, are intrinsically disordered.  At present,
there is little consensus on the effects of this disorder,
particularly near the transition to the Mott-insulating phase.

Here, we discuss the effects of strong electronic
correlations on disordered two-dimensional (2D) materials via a
numerical study of the Anderson-Hubbard model,
\begin{equation}
\hat H = -t \sum_{\langle i,j \rangle}\sum_\sigma c^\dagger_{i\sigma}
c_{j\sigma} + \sum_i \left ( U \hat n_{i\uparrow}\hat n_{i\downarrow}
+ \epsilon_i \hat n_i \right ),
\label{ham}
\end{equation}
where $\langle i,j\rangle$ refers to nearest neighbor lattice sites
$i$ and $j$, $\sigma =\uparrow,\downarrow$ is the spin index,
and $\hat n_i = \hat n_{i\uparrow} + \hat n_{i\downarrow}$ where
$\hat n_{i\sigma} = c^\dagger_{i\sigma}c_{i\sigma}$ is the local
charge density operator.  The model
has four parameters: the kinetic energy $t$,
the intra-orbital Coulomb interaction $U$, the width $W$ of the
disorder-potential distribution, and the chemical potential $\mu$.  
Disorder is introduced through randomly chosen site energies
$\epsilon_i$, which in this work are box-distributed according to
$P(\epsilon_i) = W^{-1} \Theta(W/2 - |\epsilon_i|)$.

For $U=0$, it is well understood that the single-particle eigenstates
of Eq.~(\ref{ham}) are Anderson localized for $W>W_c$, where $W_c$ is the
critical disorder and $W_c=0$ in two and fewer dimensions.  For $W=0$,
and at half-filling (ie.\ $n=1$, where $n$ is the charge density), there
is a critical interaction strength $U_c$ such that the model is a
gapped Mott insulator for $U>U_c$ and (neglecting possible
broken-symmetry phases) a strongly-correlated metal for $U<U_c$ or for
$n\neq 1$.  
There is evidence that the Mott transition is fundamentally different
in the presence of disorder.  For example, some work has shown that
$U_c = 0$ in clean low-dimensional systems with nested Fermi
surfaces,\cite{Lieb1968,Hirsch1985,White1989,Otsuka2000} while in the
disordered case $U_c$ is not only nonzero, but rapidly becomes of
order the bandwidth as a function of
$W$.\cite{Otsuka2000,Otsuka1998,Tusch1993}
In fact, it has become clear through numerous studies that the general
$U$-$W$-$n$ phase diagram is complicated and also potentially contains
superconducting, antiferromagnetic, and spin glass
phases.\cite{Tusch1993,Heidarian2003,Fazileh2006,Ulmke1995}  Much of
the recent progress has been for infinite-dimensional
systems\cite{Ulmke1995,Vollhardt2005,Balzer2005,Laad2001,Lombardo2006,Tanaskovic2003,Aguiar2005,Aguiar2007}
and the applicability of this work to two and three dimensions is not
well established.  A recent focus has been the extent to which
interactions screen the disorder
potential\cite{Srinivasan2003,Tanaskovic2003,Chakraborty2006,Aguiar2007}
and whether screening may lead to an insulator-metal
transition.\cite{Kotlyar2001,Srinivasan2003,Heidarian2003,Chakraborty2006}
In particular, some calculations show perfect screening near the Mott
transition.\cite{Tanaskovic2003,Aguiar2007}

Part of the confusion surrounding the Anderson-Hubbard phase diagram
stems from the variety of theoretical approaches that have been
applied.  Self-consistent Hartree-Fock (HF)
calculations\cite{Tusch1993,Heidarian2003,Fazileh2006} treat the
disorder potential exactly but do not capture the strong-correlation
physics of the Mott transition.  Dynamical mean-field theory (DMFT)
approaches contain the necessary strong-correlation physics but are
based on a local approximation that generally precludes exact
treatment of disorder.  A variety of coherent-potential-approximation
(CPA) and CPA-like approximations have been employed in conjunction
with
DMFT.\cite{Ulmke1995,Laad2001,Vollhardt2005,Balzer2005,Tanaskovic2003,Aguiar2005,Lombardo2006}
While the CPA reproduces some disorder-averaged quantities accurately,
eg.\ the density of states (DOS) in the noninteracting limit, it fails
to reproduce quantities that depend on explicit knowledge of spatial
correlations between lattice sites.  As a notable example, the CPA
fails to predict the Altshuler-Aronov DOS anomaly which appears at the
Fermi energy in disordered metals.\cite{Altshuler1985}  Quantum Monte
Carlo\cite{Srinivasan2003,Chakraborty2006} and exact
diagonalization\cite{Kotlyar2001} (ED) methods treat both disorder and
interactions exactly, but suffer from severe finite-size limitations,
typically generate only equal-time correlations, and (in the case of
quantum Monte Carlo) suffer from the fermion sign problem.

In this work, we use an extension of DMFT, known as statistical
DMFT,\cite{Kotliar1997} that incorporates both strong correlations and
an exact treatment of the disorder potential.  By varying $U$ for a
fixed disorder strength, we are able to move smoothly from the
well-understood weakly-correlated regime into the unknown territory of
strongly-correlated disordered systems.  Because our approach retains
spatial correlations between the local self-energy at different
lattice sites, but can also be applied to reasonably large lattices in
finite dimensions, it provides a bridge between the various methods
described above.  Up to now, statistical DMFT has been applied only on
a Bethe lattice.  Here, we work with a two-dimensional square lattice
and present results for the density of states, scaling of the inverse
participation ratio, and the screened potential.

\section{Method}
  We focus on paramagnetic solutions on the 2D square
lattice, for which the noninteracting bandwidth is $D=8t$.  On an
$N$-site lattice, the single-particle Green's function can be
expressed as an $N\times N$ matrix in the site-index:
\begin{equation}
{\bf G}(\omega) = [\omega{ \bf I} - {\bf t} - {\boldsymbol \epsilon} - {\bf
\Sigma}(\omega)]^{-1}
\label{G}
\end{equation}
with ${\bf I}$ the identity matrix, ${\bf t}$ the matrix of hopping
amplitudes, ${\boldsymbol \epsilon}$ the diagonal matrix of site
energies $\epsilon_i$ and ${\bf \Sigma}(\omega)$ the matrix of
self-energies.  The matrix ${\bf t}$ has nonzero matrix elements
$t_{ij} = -t$ for $i$ and $j$ corresponding to nearest-neighbor sites.
We assume that ${\bf \Sigma}(\omega)$ is local, having 
only diagonal matrix elements $\Sigma_i(\omega)$.

The iteration cycle begins with the calculation of ${\bf G}(\omega)$
from Eq.~(\ref{G}). For each site $i$, one defines a Weiss mean field
${\cal G}^0_i(\omega) = [G_{ii}(\omega)^{-1} + \Sigma_i(\omega)]^{-1}$
where $G_{ij}(\omega)$ are the matrix elements of ${\bf G}(\omega)$.
As in the conventional DMFT, one then solves for the full Green's
function ${\cal G}_i(\omega)$ of an Anderson impurity
whose noninteracting Green's function is ${\cal G}^0_i(\omega)$.  The
self-energy is then updated according to
$\Sigma_i^\textit{new}(\omega) = {\cal G}^0_i(\omega)^{-1} - {\cal
G}_i(\omega)^{-1}$ and the iteration cycle is restarted.  In the
disorder-free case, this algorithm reduces to the conventional DMFT.

We have used the Hubbard-I (HI) approximation as a solver for ${\cal
G}_i(\omega)$, for which, in the paramagnetic case, ${\cal G}_i(\omega)
=\left[ {\cal G}^0_i(\omega)^{-1} - \Sigma_{i}^{HI}(\omega) \right
]^{-1}$ where
 \begin{equation}
 \Sigma_{i}^{HI}(\omega) = U\frac{n_i}{2} + \frac{U^2\frac {n_i}{2} (1-\frac {n_i}{2})}{\omega - \epsilon_i
 -U(1-\frac {n_i}{2})},
\label{eq:SE}
 \end{equation}
and $n_i\equiv \langle \hat n_i\rangle$ is self-consistently determined for each site.  The HI
approximation is the simplest improvement over Hartree-Fock (HF) that
generates both upper and lower Hubbard bands. When either the HF or HI
approximations are used as a solver, the DMFT result is in fact
equivalent to the result obtained directly from the corresponding
approximation.  However, the statistical-DMFT procedure outlined
above, applied to a physical lattice, and hence implemented through a
matrix inversion [Eq.~(\ref{G})], is very computationally intensive.
In particular, it can be difficult to achieve a converged
self-consistent solution to the $N$ coupled equations for
$\Sigma_i(\omega)$.  This work, aside from being a simple improvement
beyond HF, represents an important proof of principle regarding the
feasibility of this statistical DMFT approach.

The fact that we are studying disordered systems bears on two
important and related issues: (i) the accuracy of the HI approximation
and (ii) the validity of single-site 
(as opposed to cluster or cellular)
DMFT in a finite dimensional
system.  The strengths and weaknesses of the HI approximation are well
documented in the clean limit: it is exact in the atomic limit ($U/t \gg
1$) but is nonconserving and fails to satisfy Luttinger's
theorem.\cite{Gebhard1997}  However, it is uniquely effective in
low-dimensional disordered systems because $W/t\gg 1$ corresponds to
the atomic limit for arbitrary $U$ in systems with finite coordination
number.  This is in apparent contradiction to the general result that
DMFT is only exact in infinite dimensions, while nonlocal terms in
the self-energy, neglected in DMFT, can have a dramatic effect on the
Mott transition in 2D.\cite{Moukouri2001,Imada2007} In the disordered case,
however, we have compared our results with ED studies on small
clusters\cite{Sinan2007} and have found qualitative agreement for
$W=12t$ (ie.\ $W=1.5D$), which is the focus of the current work.  By
pushing the system towards the atomic limit, strong disorder both enhances
the accuracy of the HI approximation and also reduces the importance
of nonlocal terms in the self-energy.


\begin{figure}[tb]
\includegraphics[width=\columnwidth]{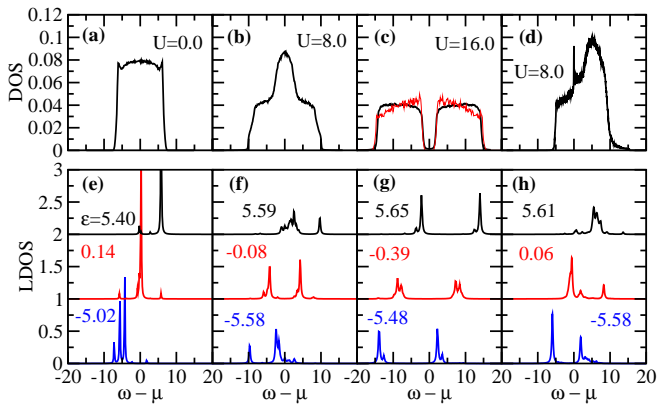}
\caption{(color online) Density of States.  (a)-(c) Total DOS 
 for $W=12t$, $n=1$ and different $U$.
Results have been averaged over 1000 samples and are for an
$N=20\times 20$ site square lattice.  
ED results, averaged over 10000 samples, for a
4-site lattice (red) are also shown in (c).
(d) DOS for $n=0.5$.
Also shown (e)-(g) is the LDOS
for three arbitrarily-chosen sites (site energies indicated in the figure).
Panel (e) is for the same parameters as (a), (f) as (b), etc.
Energies are in units of $t$.}
\label{fig:DOS}
\end{figure}

\section{Results}
  The local density of states (LDOS) is extracted from the Green's
function as $\rho({\bf r}_i,\omega) = -\pi^{-1}\mbox{Im}
G_{ii}(\omega)$ and the density of states (DOS) is $\rho(\omega) =
N^{-1} \sum_i \rho({\bf r}_i,\omega)$.  Figures~\ref{fig:DOS}(a)-(c)
show the evolution of the DOS as a function of $U$ for fixed $W$ at
half-filling.  There is a transition from a single band to a gapped
Mott-insulating state at a critical interaction $U_c\approx 12.5t$.  In
contrast, the LDOS [Figs.~\ref{fig:DOS}(e)-(g)] develops a local Mott
gap at much smaller values of $U$.  We note that, at most sites, the
noninteracting LDOS has a single strong resonance near the bare site
energy [Fig.~\ref{fig:DOS}(e)] indicating proximity to the atomic
limit where HI becomes exact.
The coordination number of the
lattice determines the disorder strength required to approach the
atomic limit.  In lattices with infinite coordination number, each
site couples to a continuum of states and the atomic limit is never
reached for finite $W$.  Thus, somewhat paradoxically, DMFT with
the HI solver works best, in disordered systems, for lattices with low coordination number.

It is useful to compare the DOS evolution in Fig.~\ref{fig:DOS} with
existing published work.  First, we note that the Mott transition
occurs at $U_c\approx W$ in both our DMFT and ED calculations.  This
is consistent with QMC results for three dimensions,\cite{Otsuka2000}
as well as infinite-dimensional DMFT results,\cite{Vollhardt2005} but
$U_c$ is somewhat larger than found in one dimension.\cite{Otsuka1998}
Our $U_c$ is also significantly larger than in unrestricted HF
calculations for three dimensions;\cite{Tusch1993} however, in HF
calculations, the metal-insulator transition is of the Slater-type and
could therefore be expected to respond differently to disorder.

Previously published results for the DOS near the Mott
transition in the large-disorder limit, to our knowledge, are for
infinite dimensions where CPA-like approximations can be made.  The
main distinction between our results and those for infinite-dimensions
is the quasiparticle resonance due to Kondo screening of the local
moments that appears at the Fermi energy for $U \sim U_c$ in the
latter
case.\cite{Tanaskovic2003,Vollhardt2005,Aguiar2005,Lombardo2006} (The
peak at $\omega = \mu$ for $U=8t$ comes from the overlap of the lower
and upper Hubbard bands.)  The HI solver cannot give such a peak;
however, ED calculations for small clusters\cite{Sinan2007} also find
no resonance at any $U$.  In particular, Fig.~\ref{fig:DOS}(c) shows
ED results for parameters corresponding to a maximum quasiparticle
resonance height in Refs.~[\onlinecite{Vollhardt2005,Aguiar2005}].
The difference between our results and those for infinite-dimensions 
might seem unsurprising since there is also no
resonance in the clean limit in 2D.\cite{Imada2007} However, the
reason for this absence appears to be different in the clean and
disordered systems. The lack of a quasiparticle resonance in clean 2D
systems has been attributed to nonlocal terms in the
self-energy,\cite{Imada2007} terms which are not included in HI.
These nonlocal terms describe short-ranged antiferromagnetic
correlations whose effect is to suppress $U_c$ to zero in the clean
limit.\cite{Hirsch1985,White1989,Moukouri2001} In our work, the absence of a quasiparticle
peak stems from the calculations being in the atomic limit.  That the same physics
controls the ED results is supported by the fact that $U_c$ is nearly
identical in HI and ED calculations.  It therefore appears as if the
limitations of the HI solver do not hide key physics in the large
disorder limit.


The DOS at quarter-filling, Fig.~\ref{fig:DOS}(d), {\em does} have a
peak at the Fermi energy, but the origin of this peak is unrelated to
strong correlations. Paramagnetic HF calculations\cite{Altshuler1985}
have shown that nonlocal charge-density correlations lead to a
positive DOS anomaly at the Fermi level in disordered metals when the
interaction is zero-range and repulsive.
This Altshuler-Aronov DOS
anomaly is absent at half-filling in our calculations, in
contradiction with the HF calculations, indicating that strong
correlations suppress the peak.  We will discuss this point below.

\begin{figure}[tb]
\includegraphics[width=\columnwidth]{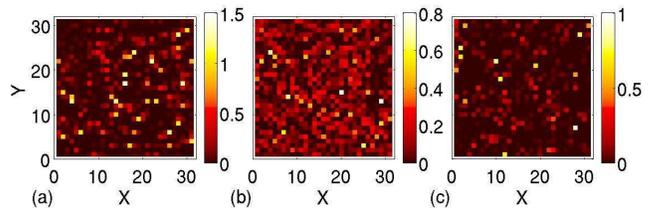}
\caption{(color online) Local density of states for a single disorder
realisation at $\omega=\mu$, $W=12t$, for (a) $U=0$, (b) $U=8t$, (c)
$U=12t$ for an $N=32\times 32$ site lattice and $n=1$.}
\label{fig:LDOS}
\end{figure}

In Fig.~\ref{fig:LDOS}, $\rho({\bf r},\mu)$ is plotted for different
values of $U$ 
for a particular disorder configuration.  
At $U=0$,
sites which have significant spectral weight at $\omega=\mu$ are
typically isolated from one another, consistent with electrons being
Anderson-localized.  The LDOS is more homogeneous for $U=8t$ than for
$U=0$, consistent with an interaction-driven delocalizing effect.
However, at $U=12t$ the LDOS is again highly inhomogeneous.

\begin{figure}[tb]
\includegraphics[width=\columnwidth]{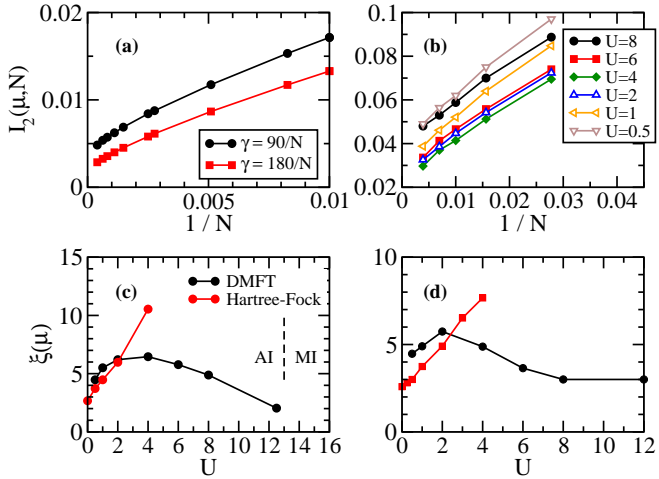}
\caption{(color online) IPR scaling with system size for
$W=12t$. Shown are (a) the dependence of the IPR on $\gamma$ for
$U=0$, (b) the scaling of the IPR with $\gamma = 4t/N$ and $n=1$,
and the dependence of the localization length (defined in text) on
$U$ at (c) $n=1$ and (d) $n=0.5$.  All curves are for $\omega = \mu$.
Anderson-insulating (AI) and Mott-insulating (MI) phases are indicated
in (c).  }
\label{fig:IPR}
\end{figure}

The inverse participation ratio (IPR)
\begin{equation}
I_2(\omega,N) = \frac{\sum_{i=1}^N \rho({\bf r}_i,\omega)^2}
{\left [\sum_{i=1}^{N} \rho({\bf r}_i,\omega)\right]^2}.
\end{equation}
provides a quantitative measure of the inhomogeneity of $\rho({\bf
r},\omega)$.  The IPR can also be used to distinguish extended and
localized states since $\lim_{N\rightarrow \infty} I_2(\omega,N) =0$
for the former and is nonzero for the latter.  It is important to note
that the frequency $\omega$ used in the calculation of ${\bf
G}(\omega)$ in Eq.~(\ref{G}) contains, by necessity, a small imaginary
component $i\gamma$.  Scaling quantities such as the IPR that are
derived from ${\bf G}(\omega)$ generally depend on
$\gamma$.\cite{Song2007}  In our scaling calculations, we have taken
$\gamma\propto 1/N$ such that the ratio of $\gamma$ to the level
spacing remains constant.  As shown in Fig.~\ref{fig:IPR}(a) for the
noninteracting case, the effect of $\gamma$ is to reduce
$I_2(\omega,N)$, such that our results represent a lower bound on the
true ($\gamma=0$) IPR.
The IPR scaling at half-filling is shown in Fig.~\ref{fig:IPR}(b) for
$\omega = \mu$.  For each value of $U$, we extrapolate a limiting
value $I_2(\mu,\infty)$.  We then define a localization length $\xi =
I_2(\mu,\infty)^{-1/d}$, where $d$ is the dimension of the
system.\cite{Mirlin1992}  For finite $\gamma$,
$I_2(\mu,\infty)^{-1/d}$ gives an upper bound for $\xi$.  

Figures~\ref{fig:IPR}(c) and (d) illustrate the effect of strong
correlations on localization. While 
the HI results
 agree closely with
self-consistent HF calculations for small $U$, the discrepancy between
the methods grows as $U$ is increased.  In the weakly-correlated small-$U$
regime, $\xi$ grows with $U$, consistent with increased screening of
the impurity potential.\cite{Tanaskovic2003,Chakraborty2006} For large
$U$, however, $\xi$ is a {\em decreasing} function of $U$ and is
smaller near the Mott transition than in the $U=0$ case.  As we
discuss below, this can be partially attributed to a decrease in
screening due to strong correlations, although screening no longer
provides a complete framework for understanding the evolution of
$\xi$.

\begin{figure}[tb]
\begin{center}
\includegraphics[width=\columnwidth]{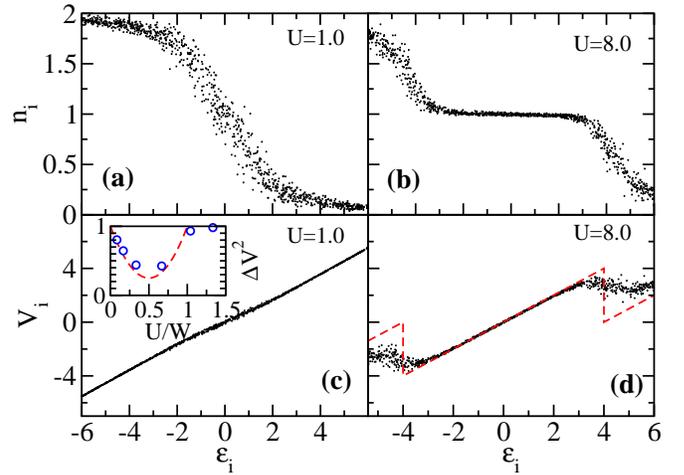}
\caption{(color online)
(a),(b) Local charge density 
and (c),(d) corresponding screened potential 
as a function of site energy.
The screened potential in the $t=0$ limit (red dashed) is also
shown in (d).  The relative variance of $V_i$ (circles) is
shown (inset) along with the $t=0$ result (dashed curve).}
\label{fig:screen}
\end{center}
\end{figure}

To understand better the localizing effect of strong correlations, we
define a screened
potential\cite{Tanaskovic2003,Chakraborty2006,Aguiar2005} $V_i$ based
on the HF site-energy: $V_i = \epsilon_i +
U\frac{n_i-n}{2}$.\cite{Chakraborty2006}  Plots of both $n_i$ and
$V_i$ as a function of $\epsilon_i$ are shown in
Fig.~\ref{fig:screen}.  In the weakly-correlated limit ($U=t$), $n_i$
is approximately linear in $\epsilon_i$ over a wide range and $V_i
\approx \epsilon_i (1 - U \chi_{ii} )$ with $\chi_{ii} = -d
n_i/d\epsilon_i$.  Since $\chi_{ii}$ depends only weakly on $U$, $V_i$
is a decreasing function of $U$.  In the strongly-correlated limit
($U=8t$), $n_i$ is a nonlinear function of $\epsilon_i$.  For
$|\epsilon_i| < U/2$, $n_i\approx 1$ and $\chi_{ii}=0$, such that
these sites are unscreened (ie.\ $V_i = \epsilon_i$).  There are,
therefore, two limits in which screening is small at half-filling:
$U\ll W$ where the interaction is too weak to screen the impurity
potential, and $U \sim W$ where strong correlations enforce
single-occupancy of most sites.
We note that the absence of an Altshuler-Aronov DOS anomaly near
half-filling can be understood in this context: strong correlations
suppress the local response of the charge density to the impurity potential.

A measure of the screening is given by the relative variance $\Delta
V^2$, defined as the variance of $V_i$ divided by the variance of
$\epsilon_i$.  Our numerical results, Fig.~\ref{fig:screen}, show that
$\Delta V^2 $ is a nonmonotonic function of $U$ obtaining a minimum at
$U\approx W/2$ and approaching $\Delta V^2 = 1$ for $U\rightarrow 0$
and $U \gtrsim W$.  At a qualitative level, this is consistent with
the nonmonotonic dependence of $\xi$ on $U$, since one expects $\xi$
to be large when $ \Delta V^2 $ is small.  There are, however,
quantitative discrepancies which show that $ \Delta V^2$ does not tell
the whole story.  First, $\xi$ does not obtain its maximum at $U=W/2$,
where $\Delta V^2$ obtains its minimum.  Second, $\xi$ is smaller at
large $U$ than at $U=0$, indicating that states near the Mott
transition are more strongly localized than at $U=0$.

Our results for the IPR are consistent with recent quantum Monte Carlo
calculations of the dc conductivity\cite{Kotlyar2001,Chakraborty2006}
which find a similar nonmonotonic dependence on $U$.
The results for the screened potential, however, are inconsistent with
infinite-dimensional DMFT results.\cite{Tanaskovic2003,Aguiar2005} In
Ref.~[\onlinecite{Tanaskovic2003}], impurities are found to be perfectly
screened (ie.\ $\Delta V^2 \rightarrow 0$) at the Mott transition,
which would correspond to a divergent $\xi$ in our calculations.  Near
the Mott transition, Ref.~[\onlinecite{Aguiar2007}] found that sites with
$|\epsilon_i| < U/2$ are perfectly screened.  Both of these results
are opposite to what we have found here.

There are two important distinctions between our calculation and those
of Refs.~[\onlinecite{Tanaskovic2003,Aguiar2007}],
both of which contribute to these opposing results on
screening.  First, Refs.~[\onlinecite{Tanaskovic2003,Aguiar2007}]  use an effective medium approach for
the disorder potential that results in metallic behavior for small U.
This is reasonable in high dimensions where Anderson localization only
occurs for strong disorder.  Our exact treatment of the disorder
potential, on the other hand, allows us to describe the Anderson
localized phase that occurs in 2D for small U.  Whereas the LDOS is
continuous and relatively uniform in the metallic case, it is
inhomogeneous and dominated by small numbers of resonances in the
Anderson insulating phase [cf.\ Fig. \ref{fig:DOS}(f)].  The local charge
susceptibility $\chi_{ii}$ is suppressed for sites with small LDOS at the
Fermi level, and screening in the Anderson-insulating phase is
consequently expected to be less than in the metallic phase.  The second
distinction is the quasiparticle resonance which arises in the
infinite-dimensional case, but does not occur in our calculations.
This resonance is a key factor in the perfect screening found in
Refs.~[\onlinecite{Tanaskovic2003,Aguiar2007}].

In conclusion, we have studied the 2D Anderson-Hubbard model at half-
and quarter-filling, in the limit of large disorder using statistical
DMFT.  We have calculated the localization length $\xi$ from
the inverse participation ratio, and find that it varies
nonmonotonically with the strength of the interaction: at small $U$,
the interaction screens the impurity potential, but at large $U$
strong correlations reduce the screening.  As a consequence, the
Altshuler-Aronov DOS anomaly is suppressed at half-filling.
For strong disorder, we find no evidence for an insulator-metal transition nor
for enhanced screening near the Mott transition.  

\section*{Acknowledgments}
 We thank R. J. Gooding and E. Miranda for
helpful conversations.  We acknowledge Trent University, NSERC of
Canada, CFI, and OIT for financial support.  Some calculations were
performed using the High Performance Computing Virtual Laboratory
(HPCVL).

\end{document}